# Empowering ChatGPT-Like Large-Scale Language Models with Local Knowledge Base for Industrial Prognostics and Health Management

Huan Wang and Yan-Fu Li, *IEEE Senior Member*, and Min Xie, *IEEE Fellow*

*Abstract*—Prognostics and health management (PHM) is essential for industrial operation and maintenance, focusing on predicting, diagnosing, and managing the health status of industrial systems. The emergence of the ChatGPT-Like large-scale language model (LLM) has begun to lead a new round of innovation in the AI field. It has extensively promoted the level of intelligence in various fields. Therefore, it is also expected further to change the application paradigm in industrial PHM and promote PHM to become intelligent. Although ChatGPT-Like LLMs have rich knowledge reserves and powerful language understanding and generation capabilities, they lack domain-specific expertise, significantly limiting their practicability in PHM applications. To this end, this study explores the ChatGPT-Like LLM empowered by the local knowledge base (LKB) in industrial PHM to solve the above limitations. In addition, we introduce the method and steps of combining the LKB with LLMs, including LKB preparation, LKB vectorization, prompt engineering, etc. Experimental analysis of real cases shows that combining the LKB with ChatGPT-Like LLM can significantly improve its performance and make ChatGPT-Like LLMs more accurate, relevant, and able to provide more insightful information. This can promote the development of ChatGPT-Like LLMs in industrial PHM and promote their efficiency and quality.

## I. Introduction

PROGNOSTICS and health management (PHM) constitutes an indispensable component within contemporary industrial operation and maintenance practices, focusing on the prediction, diagnosis, and management of the well-being of industrial equipment and systems [1, 2]. Complex system degradation models provide a strong guarantee for the reliability and safety of industrial equipment [3, 4]. By promptly detecting and prognosticating the health status, PHM contributes to the prevention of unforeseen equipment shutdowns, cost reduction in maintenance activities, and enhancement of production efficiency and product quality. Given the escalating complexity and diversification of modern industrial equipment, a considerable workforce of proficient operation and maintenance personnel is essential for analyzing, processing, repairing, and replacing diverse industrial data and equipment. Furthermore, the aforementioned procedures demand the expertise of experienced engineers to ensure the secure operation of industrial equipment [5]. Additionally, with the proliferation of industrial sensors, a vast volume of operational data is collected, rendering sole reliance on manual processing by operation and maintenance engineers impractical. Consequently, the development of automated systems for PHM data analysis and processing is paramount in industrial operation and maintenance.

Over the past decade, the advancement of deep learning technology has facilitated the automation of industrial PHM data analysis and feature extraction [6-8]. Deep learning strives to address intricate tasks by constructing and simulating artificial neural networks [9]. At its core, deep learning endeavors to extract abstract features from data through multi-level nonlinear network layers, enabling automated pattern recognition and decision-making. Consequently, deep learning has exhibited remarkable achievements across numerous domains [10, 11], encompassing computer vision [12], natural language processing (NLP) [13], and speech recognition. Notably, its extraordinary accomplishments extend to the domain of PHM as well. As a result, deep learning finds extensive employment in diverse PHM applications, such as fault prediction [14-16], anomaly detection [17, 18], and maintenance planning [19], to realize automated data analysis, intelligent prediction, and decision-making. For example, the Semi-Supervised deep SVDD algorithm proposed by Peng et al. [18] effectively solves the data scarcity problem and can achieve early fault warning. Wang et al. [20] proposed a Transformer-based train wheel wear prediction model, which can provide valuable wheel health status information for maintenance personnel. Han et al. [21] proposed a deep model-based domain adaptation algorithm and realized lithium-ion battery capacity estimation under invisible conditions.

The significance of deep learning in facilitating automatic data analysis and intelligent decision-making within industrial PHM is evident [22, 23]. It empowers operation and maintenance personnel with timely and valuable information, thereby enhancing the maintenance standards of industrial equipment in enterprises. However, the scope of deep learning's application remains limited, excelling primarily in data analysis and certain monitoring and forecasting applications. Consequently, the current operation and maintenance of industrial equipment and systems necessitate a

This work was supported by the National Natural Science Foundation of China (71731008) and the Beijing Municipal Natural Science Foundation-Rail Transit Joint Research Program (L191022). (Corresponding author: Yan-Fu Li, liyanfu@tsinghua.edu.cn).

Huan Wang and Yan-Fu Li are with the Department of Industrial Engineering, Tsinghua University, Beijing, 100084, China.
Min Xie is with the department of Advanced Design and System Engineering, City University of Hong Kong, Hong Kong SAR, China.

substantial workforce comprising engineers possessing extensive professional knowledge to conduct monitoring, troubleshooting, equipment maintenance, and related tasks, ensuring operational safety and reliability [24]. Given the complexity and specialization of modern industrial equipment and systems, these roles demand a high degree of specialized expertise and experience. The challenge at hand for industrial enterprises lies in swiftly training engineers with professional knowledge and reducing the labor costs associated with industrial equipment monitoring, operation, and maintenance- an issue that cannot be disregarded.

In the past two years, deep learning has witnessed revolutionary breakthroughs in the domain of NLP, with the gradual maturation of large-scale language models (LLM) [25, 26] that possess formidable language understanding capabilities and the ability to engage in human-like communication. LLMs leverage extensive text training data and intricate network models to discern the underlying patterns and principles of language, enabling them to generate accurate, contextually appropriate, and fluent responses based on the input. Among the groundbreaking achievements of LLMs, ChatGPT [27] emerged as a global sensation immediately upon its release. Its vast knowledge base, intelligent responsiveness, and flawless answers instilled hope in the realm of general artificial intelligence. Consequently, ChatGPT-like LLMs harbor the revolutionary potential for applications in various fields [28]. For instance, ChatGPT-like LLMs can supplant product customer service, intelligently and fluently providing professional responses to customer inquiries. They can assist physicians in medical image detection, enabling the identification of disease types and severity while automatically generating professional medical reports [29]. Moreover, ChatGPT-like LLMs can utilize their powerful summarization capabilities to enhance search engines, furnishing users with more precise and pertinent content.

ChatGPT-Like LLM holds immense potential to revolutionize the application paradigm in PHM, facilitating the development of intelligent industrial operation and maintenance systems while concurrently reducing labor costs associated with industrial equipment monitoring, operation, and maintenance [30]. For instance, ChatGPT-Like LLM can serve as a comprehensive knowledge base, delivering reliable and insightful information to operation and maintenance personnel by elucidating complex issues and furnishing operational guidelines. LLM can analyze equipment operation data to generate analysis reports and visualizations, aiding decision-making. Integration of LLM into industrial automation systems can bestow intelligent automated workflows. Furthermore, LLM can provide expert repair advice and guidance, minimizing unplanned downtime and mitigating maintenance expenses. Furthermore, ChatGPT-Like LLM holds vast application prospects within the realm of industrial PHM, promising to reshape the entire field's working paradigm in the future.

However, applying the current ChatGPT-Like LLM to industrial PHM entails several challenges and limitations. While these language models possess formidable language processing capabilities and vast knowledge reserves, a majority of their training data originates from the Internet. Consequently, technical content, technical manuals, and maintenance guides prevalent in the industrial domain are often confidential and inaccessible. As a result, existing LLMs often lack expertise in industrial PHM, rendering them unable to accurately comprehend and address domain-specific concerns related to industrial equipment monitoring, troubleshooting, and maintenance. Consequently, practical application encounters difficulties such as misconceptions and the delivery of inaccurate or irrelevant information in response content.

To address these limitations, a promising solution involves integrating a local knowledge base (LKB) with ChatGPT-Like LLM to enhance the model's expertise and practicality. In this study, we delve into this solution, thoroughly examining the methods and steps involved in leveraging the LKB to empower ChatGPT-Like LLM. These include knowledge base preparation, text vectorization, similarity search, and language model integration. We present two real-world cases as experimental evidence to validate the practicality and performance improvement of ChatGPT-Like LLM empowered by the LKB (LKB-E-LLM) in industrial PHM. Through comprehensive experimental analysis and result evaluation, we showcase the significant impact of integrating LKBs on language models, emphasizing their advantages in fault diagnosis, prediction, and maintenance management. Additionally, we discuss potential applications and benefits of employing LKBs while outlining directions for future research. The contributions of this study can be summarized as follows:

1. This paper investigates the application potential of ChatGPT-Like LLM in the industrial PHM, exploring its current limitations and proposing potential solutions.
2. This paper examines the integration of the LKB with ChatGPT-Like LLM in the context of industrial PHM. The study demonstrates how this integration significantly enhances the performance and practicality of LLM in PHM, thereby facilitating its real-world application.
3. This paper empirically validates the practicality and performance improvements of LKB-E-LLM in PHM through actual cases, highlighting its advantages in fault diagnosis, prediction, and maintenance management.
4. This paper provides a comprehensive analysis and discussion of the application potential of LKB-E-LLM in the industrial PHM, while also exploring future research directions in this area.

The rest of this paper is organized as follows. Section II introduces the key components of ChatGPT-Like LLM. Section III describes the technical details of LKB-E-LLM. Section IV verifies the performance of LKB-E-LLM. Section V discusses the advantages, potential applications, and future directions of LKB-E-LLM. Section VI summarizes this paper.

## II. ChatGPT-Like Large-Scale Language Model

LLM is a language model characterized by extensive parameters, advanced semantic comprehension, and text generation capabilities [31]. Through extensive training on vast textual data, LLM captures language's intricacies and fundamental characteristics, thereby facilitating the generation of high-quality natural language text and enabling effective communication with human counterparts. Many LLMs are constructed based on diverse pre-trained language models, further enhanced by techniques such as supervised fine-tuning, feedback bootstrap, and reinforcement learning from human feedback [27]. These techniques enable the model to discern the intended meaning of human instructions and produce suitable responses. An illustrative example of an LLM is the ChatGLM-6B model [32] developed by Tsinghua University. ChatGLM-6B is a Chinese-English bilingual language model with 6.2 billion parameters. To train ChatGLM-6B, an autoregressive blank infilling approach is employed as the core self-supervised training method, leveraging extensive Chinese and English text datasets. Additionally, the ChatGLM-6B incorporates 2D positional encodings and enables the prediction of spans in arbitrary order. These enhancements contribute to more effective self-supervised pre-training and elevate the model's performance in natural language understanding applications [33]. When provided with an input text $X=[x_1, x_2, \cdots, x_n]$, it samples multiple text spans $\{S_1, S_2, \cdots, S_m\}$, with each span $S_i$ representing a sequence of consecutive tokens $[s_{i,1}, s_{i,2}, \cdots, s_{i,l_i}]$ in $X$. To create a corrupted text $X_{\text{corrupt}}$, each span is substituted with a single [MASK] token. The model utilizes the autoregressive approach to predict the missing tokens within the spans based on the corrupted text. This means that while predicting the missing tokens in a span, the model can access the corrupted text and the previously predicted spans. In order to effectively capture the interdependencies among different spans, this method randomly shuffles the order of the spans. Let $Z_m$ represent the set of all possible permutations of an index sequence [1, 2, ..., m] with length $m$. Given a span $S_{z<i}=[S_{z_1}, \cdots, S_{z_{i-1}}]$, The pre-training objective is then defined as:

$$\mathcal{L} = \max_{\vartheta} \mathbb{E}_{z \sim Z_m} \left[ \sum_{i=1}^{m} \log p_\vartheta (S_{z_i} | X_{corrupt},\ S_{z<i}) \right] \quad (1)$$

The tokens within each blank are always generated in a left-to-right order, whereby the probability of the span $S_i$ being generated is factorized as:

$$p_\vartheta (S_i | X_{corrupt},\ S_{z<i}) = \prod_{j=1}^{l_i} p(s_{i,j} | X_{corrupt}, S_{z<i}, S_{i<j}) \quad (2)$$

In addition, ChatGLM-6B investigates a multi-task pre-training mechanism to generate appropriate responses in language understanding applications.

In addition to ChatGLM-6B, the LLM is experiencing rapid development, with notable contributions from prominent research institutions and companies. For instance, OpenAI has made significant strides with the GPT series [27, 34] of LLMs, which have garnered substantial success and widespread adoption. Among them, GPT-3.5 [27] stands out for its formidable dialogue and language comprehension capabilities, while GPT-4 [34] represents a notable advancement by incorporating image understanding capabilities, thereby facilitating highly intelligent multi-modal information processing. Baidu's ERNIE series [35, 36] of LLMs have also demonstrated exceptional performance. These LLMs are built upon the foundation of the Transformer and its enhanced versions [13, 37], utilizing the self-attention mechanism of the Transformer as the core backbone network to achieve effective feature representation and semantic comprehension. Subsequently, autoregressive models and encoder-decoder models are employed to train and optimize the LLMs on vast amounts of unlabeled text data using self-supervised representation learning [38] algorithms. Moreover, to enhance the LLM's ability to understand and address human-specific problems, methodologies such as Reinforcement Learning from Human Feedback [27] are employed to optimize the LLM. This approach leverages reinforcement learning techniques by incorporating valuable feedback provided by human experts, thus refining the LLM's responses. Looking ahead, LLMs are poised to make comprehensive contributions across various industries, propelling societal advancements.

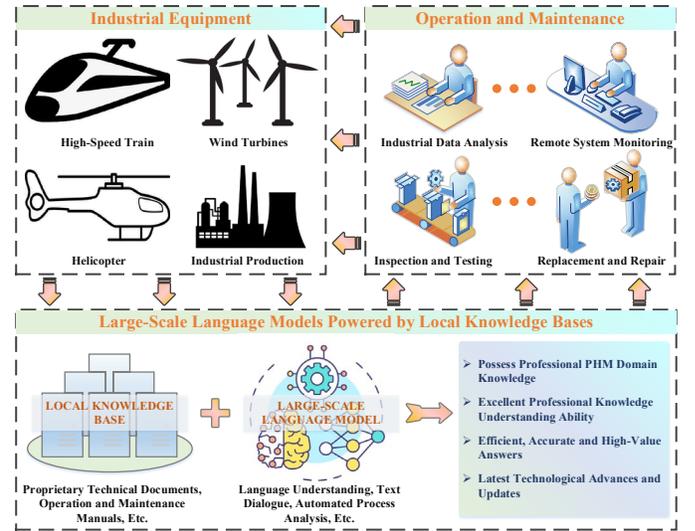

**Fig. 1** The schematic diagram of the application of LKB- empowered LLM in the industrial PHM.

## III. LLM EMPOWERED BY LOCAL KNOWLEDGE BASE

As mentioned previously, the robust language processing and semantic understanding capabilities of ChatGPT-Like LLM are poised to make a significant impact on the field of PHM, enhancing efficiency, performance, and the overall intelligence level of industrial PHM. However, the existing LLMs typically lack domain-specific expertise, such as in the case of industrial PHM, resulting in subpar responses to technical-specific queries. This limitation arises from the fact that the training data for LLMs predominantly originates from the Internet, while the crucial knowledge in industrial PHM, including technical manuals, maintenance guidelines, and performance parameters, remains confidential. Consequently, LLMs fail to acquire this essential information during the training process. Furthermore, the need to prioritize data privacy and security in the industrial PHM field prohibits the collection of private data for fine-tuning the model, as it may lead to information leakage. In **Fig. 1**, we present a promising solution that involves integrating a local private knowledge base with ChatGPT-Like LLM to equip the LLM with domain-specific expertise, thereby enhancing its practicality.

Industries such as high-speed trains, wind turbines, helicopters, and industrial production necessitate robust operation and maintenance teams to ensure equipment and system safety and reliability. The combination of ChatGPT-Like LLM and the local industrial private knowledge base holds the potential to address various aspects of the PHM field. This includes assisting in industrial data analysis to generate comprehensive analysis reports, facilitating automatic remote monitoring and early detection of abnormal situations, aiding in remote troubleshooting and equipment maintenance, and supporting personnel training and knowledge transfer. By leveraging this integrated approach, industrial enterprises can significantly reduce operation and maintenance costs while enhancing the operational safety of their equipment. The subsequent sections will delve into integrating LLM and LKB, offering a detailed overview of the integration process and its benefits.

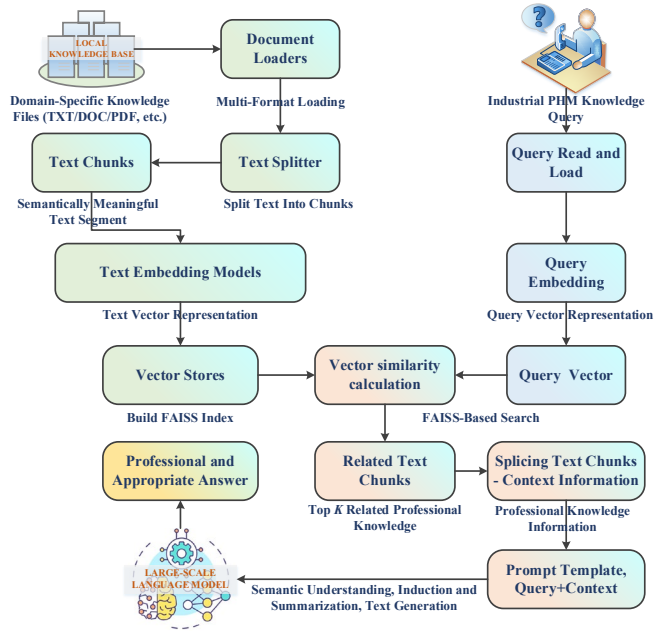

**Fig. 2** The overall system architecture of the LKB-empowered LLM.

*A. Overall System Architecture*

Fig. 2 illustrates the comprehensive system architecture of LKB-E-LLM. It encompasses several key processes, including local knowledge base construction, document reading, text splitting, text vectorization, vector similarity calculation, relevant text search, prompt content generation based on relevant text and queries, and answer generation based on prompts. The existence of abundant private technical documents related to various industrial equipment and systems enables the establishment of an offline knowledge base. Domain-specific knowledge unknown to the LLMs is gathered and stored offline to construct this knowledge base. These data are then processed by a text reader capable of handling diverse file formats such as TXT, DOC, PDF, CSV, and more.

Long text information is split into text chunks with semantic significance to enhance the processing of textual content. Text splitting involves various techniques, including character-based splitting, token-based splitting, and recursive character-based splitting. These methods aim to ensure that the split text chunks consist of semantically relevant text fragments. A text embedding model is employed to enable the model to comprehend natural language. Text embedding [39] involves converting text data into continuous vector representations. By mapping words, phrases, or sentences in the text to a continuous vector space, text embedding provides the foundation for numerous NLP tasks. The fundamental concept behind text embedding is to capture the semantic information of the text, allowing texts with similar semantics to have similar representations. Presently, text embedding methods based on large-scale pre-trained models have gained popularity due to their superior performance compared to traditional word embedding algorithms.

Lastly, to facilitate the processing of large-scale local text, the obtained vector representations need to be stored and indexed for efficient search. A highly suitable option for this purpose is Facebook AI Similarity Search (FAISS) [40]. FAISS offers a comprehensive suite of algorithms and data structures for large-scale vector retrieval. It supports parallel computing and vector quantization technology, enabling efficient search operations on vast datasets.

For queries from operation and maintenance personnel, the text reading and vector representation processes are also involved. The query text is first converted into a vector representation using the text embedding model. Subsequently, the powerful search capability of FAISS is utilized to locate the most relevant content from the LKB. The critical aspect of this process is the search and similarity calculation operation, which identifies the top $K$ relevant professional knowledge entries from the LKB. After retrieving the $K$ relevant professional knowledge entries, they are concatenated into contextual information according to a specific format. This contextual information is then combined with the query to construct the prompt input for the LLM. Ultimately, the LLM generates professional and reliable responses based on the provided information and its knowledge. Acting as the intermediary between users and the knowledge base, the ChatGPT-Like LLM utilizes its language processing and semantic understanding capabilities to analyze, summarize, and reason contextual information to produce accurate and appropriate responses.

In the above process, text embedding, vector similarity calculation, and prompt engineering are the core of LKB-E-LLM, and these key components will be introduced in detail below.

*B. Text Embedding*

Text embedding serves as the fundamental mechanism for enabling computers to comprehend natural language and establishing a bridge between human language and computational systems [39]. It involves the technique of representing textual data as continuous vectors within a low-dimensional vector space. This methodology aims to capture text's semantic and syntactic characteristics, facilitating the quantification of similarity and relationships between words, phrases, or entire documents. Notably, with the emergence of large-scale pre-training models, text embedding approaches have increasingly incorporated these models to achieve more precise and expressive embedding representations. Prominent pre-training models, such as BERT [41], GPT [27], and ERNIE [35, 36], have been widely employed in text embedding methodologies. These models offer notable advantages as they are pre-trained on extensive datasets, granting them robust capabilities for semantic and syntactic representations. Additionally, they possess the ability to generate context-aware embedding representations that effectively capture word polysemy and contextual relevance. To illustrate, taking BERT as an example, the main technical details are explained below.

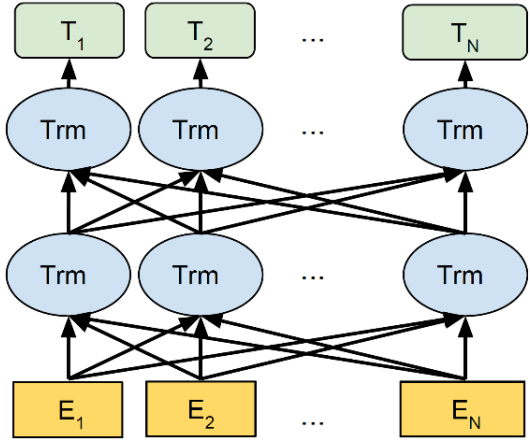

**Fig. 3** BERT's pre-training model architectures, which use a bidirectional Transformer. $E_i$ represents the input, and $T_i$ represents the output [41].

BERT [41], a Transformer-based pre-training model, leverages a bidirectional encoder to capture extensive contextual information from large-scale text, facilitating a profound understanding of text semantics and context. BERT primarily employs two methods, namely masked language model (MLM) and next sentence prediction (NSP), to achieve natural language understanding. MLM aims to encourage the model to grasp rich semantic and contextual information by selectively masking certain words within the input text and prompting the model to predict these masked words. On the other hand, NSP aims to train the model in discerning whether two sentences are consecutive in the original text, thereby fostering the model's comprehension of sentence relationships and contextual coherence. **Fig. 3** shows the model architecture of BERT. The feature representation of large-scale text is mainly completed by the Transformer encoder. The Transformer encoder comprises multiple self-attention mechanisms, which effectively capture the dependencies among various positions within the input sequence. Assuming the input to the self-attention mechanism [13] is denoted as $X = [x_1, x_2, \cdots, x_k]$, it is initially encoded into query vectors ($Q$), key vectors ($K$), and value vectors ($V$) via three distinct sets of linear transformation layers. These linear transformations, represented by the learnable weights $\mathbf{W}^Q, \mathbf{W}^K, \mathbf{W}^V$, respectively, facilitate the transformation process. The self-attention mechanism can be expressed as:

$$Att(Q, K, V) = Softmax\left(\frac{(X\mathbf{W}^Q)(X\mathbf{W}^K)^T}{\sqrt{d}}\right)(X\mathbf{W}^V) \quad (3)$$

The purpose of $QK^T$ is to calculate the similarity between $Q$ and $K$, then input the scale-transformed similarity into Softmax to obtain the attention weight and perform matrix multiplication between the weight and $V$ to obtain the final output. Further, to achieve multi-level feature encoding, multi-head attention (MHA) is adopted, which employs multiple self-attention modules in parallel. These self-attention modules encode the global correlation of the input data from different perspectives, respectively, and the results are finally combined as the output of MHA. MHA can be expressed as:

$$\begin{aligned} S &= Linear_h(MHA(X)) = Concat(head^1, head^2, \cdots head^H)\mathbf{W}^H, \\ where \quad head^H &= Att(X\mathbf{W}_H^Q, X\mathbf{W}_H^K, X\mathbf{W}_H^V) \end{aligned} \quad (4)$$

where $head^H$ represents an attention head (self-attention module), $H$ represents the number of attention heads, and $Concat(\cdot)$ represents the splicing of the output results of $H$ attention heads.

*C. Vector Similarity Calculation*

Essentially, text embedding involves encoding text chunks into dense vectors that capture their semantic and contextual information. Vector similarity calculation is a crucial technology in LKB-E-LLM. In NLP, cosine similarity is commonly employed to assess the similarity between vectors and evaluate their distribution within the semantic space. Let $v$ represent the vector representation of the knowledge base's text chunks, and $w$ represent the vector representation of the query. The cosine similarity between them can be expressed as follows:

$$\cosine(v, w) = \frac{v \cdot w}{|v||w|} = \frac{\sum_{i=i}^{L} v_i w_i}{\sqrt{\sum_{i=1}^{L} v_i^2}\sqrt{\sum_{i=1}^{L} w_i^2}} \quad (5)$$

where $L$ represents the length of the vector. In the case of a relatively small knowledge base, the process involves traversing the knowledge base vector based on the query vector to identify the most relevant text segments and extract them accordingly. However, when dealing with large and complex knowledge bases, this approach incurs significant time overhead, making it impractical. To address this challenge, the introduction of FAISS in LKB-E-LLM enables efficient similarity search at a billion-scale level, leveraging the power of GPU acceleration.

FAISS is a specialized framework designed to facilitate efficient searching and clustering of dense vectors [40]. It encompasses a range of algorithms specifically designed for searching within collections of vectors with varying dimensions. The workflow of FAISS can be broadly divided into two main steps: indexing and searching/querying. In FAISS, the most commonly utilized method for index construction is based on k-means clustering. This technique involves dividing a collection of vectors into clusters and creating an inverted file for each cluster. During index construction, different cluster numbers and cluster centers can be chosen to strike a balance between search speed and accuracy. Additionally, FAISS offers various query acceleration methods, with Product Quantization being a prominent algorithm. Product quantization involves splitting a high-dimensional vector into multiple subvectors and quantizing each subvector into a discrete codebook representation. This quantization process reduces storage requirements, and during similarity search, comparisons can be performed quickly within the quantized codebook, thereby enhancing search efficiency. Furthermore, FAISS supports GPU acceleration, multi-index support, and distributed indexes' construction and search functions. These features enable FAISS to handle large-scale vector similarity search tasks efficiently.

*D. Prompt Engineering*

Prompt engineering is a strategic approach employed in text generation applications to optimize and design prompts, thereby enhancing the performance and quality of generated text by generative models. By carefully crafting and refining prompts, generative models can be guided to produce precise, consistent, and desirable output. The crux of prompt engineering lies in devising well-suited prompts for the intended task, which involve specific formatting and content requirements. In the context of LKB-E-LLM, the methodology heavily relies on prompt engineering. It constructs an appropriate prompt based on the query vector and the outcomes of similarity searches conducted on the LKB. Typically, FAISS retrieves the $K$ most relevant text segments to the query vector, where $K$ is an adjustable hyperparameter. Assuming that the retrieved text segments by FAISS are denoted as $Text_1, Text_2, \cdots, Text_K$, the LLM prompt can be formulated using the following PROMPT TEMPLATE. By doing so, ChatGPT-Like LLMs can generate more proficient and accurate content based on the provided prompt information.

---

PROMPT_TEMPLATE = "Known Information:

$\{Text_1, Text_2, \cdots, Text_K\}$,

Based on known information, please answer relevant questions concisely and professionally.

The question is: {query}."

---

Moreover, PROMPT TEMPLATE can be tailored to specific industrial fields, considering the particular scenarios encountered. This aids in enabling the model to comprehend task context, requirements, and constraints, thereby generating task-specific and expert-level content. By employing appropriate prompts in conjunction with a vast LKB, Prompt Engineering can equip LLMs with substantial domain expertise, empowering them to comprehend the intricacies and specifications of diverse PHM applications. Consequently, ChatGPT-Like LLMs can exhibit exceptional performance within the industrial PHM domain.

*E. Implementation Details*

LKB-E-LLM is an advancement of existing LLMs, and its implementation necessitates the utilization of various LLM technologies, such as text embedding and vector similarity. The LangChain framework[1] serves as a robust support system for realizing LKB-E-LLM. It is a language model application development framework that imparts two fundamental capabilities to LLMs: data perception, enabling the language model to connect with external data sources, and agency, allowing the language model to interact with its environment. Additionally, LangChain offers interfaces for multi-format document extraction, text splitting, text embedding, and vector similarity querying, providing substantial support for realizing LKB-E-LLM.

With the advancements in large-scale pre-training models, several text embedding modules like GPT [27], BERT [41], M3E[2], and ERNIE-3.0 [35] have proven reliable for text embedding in LKB-E-LLM. Considering local implementation's efficiency and resource limitations, this study adopts the M3E model. To facilitate efficient similarity querying and searching of large-scale vector data, FAISS, Redis[3], and Milvus[4] serve as essential components integrated into the LangChain framework. Notably, projects such as langchain-ChatGLM[5] and Wenda[6] have already achieved commendable performance by leveraging scalable knowledge bases for question-answering using the aforementioned technical foundation.

This study explores the application of LKB-E-LLM in the industrial PHM, building upon the previous work. The implementation of LKB-E-LLM is carried out on a workstation equipped with an Intel 10900K CPU, GeForce RTX 3090 GPU, and 32GB memory. Considering the limitations of video memory, the LMM model employed is ChatGLM-6B [32], an efficient Chinese-English bilingual dialogue language model that requires a maximum of 14GB video memory.

## IV. INDUSTRIAL CASE STUDIES

This section takes high-speed trains (HSTs) and wind turbines as examples to verify the performance of LKB-E-LLM in specific industrial fields.

*A. LKB-E-LLM for High-Speed Trains*

As a case study, this study has constructed a small LKB related to HSTs. It mainly includes professional knowledge such as "Operating and Maintenance Implementation Rules for High-Speed Train Catenary," "Operating and Maintenance Rules for High-Speed Train Traction Substations," and "Maintenance Rules for High-Speed Railway Signaling Systems." These documents mainly involve the professional knowledge of operation and maintenance management of high-speed railway catenary, traction substation, and signal equipment. Due to limitations in the training corpus, most ChatGPT-Like LLMs struggle to fully grasp the relevant technical details. Considering the space constraints, this study presents the results of two professional technical question-and-answer pairs. Then, a comparison is made between the responses of LKB-E-LLM and regular LLM to validate the performance gains brought by the LKB. The detailed question-and-answer records are shown in TABLE I.

As shown in TABLE I, LKB-E-LLM and LLM answered the minor repairs of AC and DC power supply units of HSTs

---

and the technical standards of insulators. By comparing the responses of LKB-E-LLM and LLM, it can be observed that the responses from LLM are vague, hollow, and lack specific technical details. On the other hand, the responses from LKB-E-LLM are more professional, with explicit and clear technical details, which can effectively guide operational and maintenance personnel in their work. For example, in question 1, LKB-E-LLM clearly and professionally explains the points to be noted during the minor repair process of power supply devices. This includes checking and recording the voltage and current of the battery, inspecting the battery's appearance, and cleaning the relevant equipment from dust. In contrast, LLM's response is less specific, lacks operability, and describes tasks more suited for major repairs, such as replacing specific core components. Similarly, for question 2, LKB-E-LLM provides a comprehensive and detailed explanation of the technical standards for insulator maintenance from various aspects. It contains strong guidance with technical details. In contrast, LLM's response is broad and appears correct, but it deviates significantly from the requirements of real industrial scenarios.

The main reason for these differences is that the existing LLMs lack domain-specific expertise. When faced with specific professional questions, they are unable to provide detailed, specific, and industry-compliant answers. LKB-E-LLM significantly addresses this issue by leveraging the knowledge in the LKB, allowing LLM to possess abundant professional knowledge and respond fluently to professional questions. With the enrichment of content in the LKB, LKB-E-LLM is expected to demonstrate even better performance.

On the other hand, LKB-E-LLM also has some areas where it performs less satisfactorily, such as potentially generating partially incorrect outputs and producing text with significant redundancy. For example, in question 1, there is considerable redundancy in the response from LKB-E-LLM. This is mainly because the semantic understanding and text generation capabilities of LLM still have room for improvement. This study utilizes a relatively lightweight model, ChatGLM-6B, which has certain gaps compared to models like ChatGPT. When sufficient computational resources are available, replacing ChatGLM-6B with larger models will yield better performance.

TABLE I
LKB-E-LLM AND LLM Q&A RECORDS ON HSTs RELATED QUESTIONS

| Question-1 | Please provide a detailed description of the scope and standards for minor repairs of AC and DC power supply devices. |
|---|---|
| LKB-E-LLM | (1) Measure and record each battery's terminal voltage and float charge voltage, which should comply with the specifications. This is one of the scope and standards for minor repairs of the AC/DC power supply unit. Check each battery's terminal voltage and float charge voltage to ensure they meet the specifications stated in the manual. These checks are crucial for maintaining the performance and reliability of the AC/DC power supply unit.<br>(2) Remove dust and dirt from the DC charging unit, especially from the heat sink and cooling fan. This is also one of the scope and standards for minor repairs of the AC/DC power supply unit. Removing dust and dirt from the DC charging unit ensures the proper functioning of its cooling system and prevents overheating and other potential issues.<br>(3) Inspect the appearance of the batteries, which should be intact, clean, without deformation or bulging, and with reliable wire connections. Inspect the appearance of the batteries to ensure they are in good condition, without deformation, bulging, or other damage. Additionally, check the wire connections to ensure they are secure and reliable.<br>(4) Verify the secure installation of the DC panel and cabinets, free from corrosion and dirt and with good paint coating. Inspect the secure installation of the DC panel and cabinets, ensuring they are free from corrosion or other damages. Also, check if they are coated with appropriate paint to maintain their appearance and performance.<br>(5) Check the equipment's current, voltage, insulation monitoring data, and signal display, which should be functioning properly. Inspect the current, voltage, and insulation monitoring data of the AC/DC power supply unit to ensure its normal operation. Additionally, check the signal display and warning system to ensure they are working properly and issue necessary warnings to protect equipment and personnel safety. |
| LLM | Scope of Minor Repairs: The scope of minor repairs for AC/DC power supply units typically includes the following aspects:<br>● Replacement or repair of battery packs or charging circuits.<br>● Replacement or repair of electronic components such as switches, control boards, or oscillators.<br>● Cleaning or replacement of radiators or fans to maintain effective cooling of the power supply unit.<br>● Inspection or repair of circuit connections in the power supply unit to ensure their reliability.<br>Standards: Within the scope of minor repairs for AC/DC power supply units, there are also some general standards and procedures to ensure consistency and reliability in the maintenance process:<br>● Use appropriate tools and equipment.<br>● Follow the instructions in the standard maintenance manual or operating instructions.<br>● Avoid attempting self-repairs, especially in uncertain situations.<br>● Ensure that the repair process does not cause permanent damage or failure to the circuit or equipment.<br>The scope and standards of minor repairs for AC/DC power supply units may vary depending on the equipment and manufacturer. During the repair process, it is important to be familiar with the corresponding standards and operating instructions and follow the guidance provided by the manufacturer. |
| Question-2 | Please describe in detail the technical standards for insulator maintenance. |
| LKB-E-LLM | (1) The leakage distance of the contact wire insulating components should be ≥1400mm. This standard stipulates that the leakage distance of the insulating components should be greater than 1400mm to ensure that they are not damaged by leakage during the operation of the contact wire, thereby ensuring the safe operation of the contact wire system.<br>(2) The surface of the insulator should be clean, smooth, free from dirt, intact, and without fractures. The area of porcelain enamel peeling should not exceed 300 mm2. This ensures that the insulator surface does not leak current due to wear or damage and can withstand current surges during operation.<br>(3) The porcelain part of the insulator should have good adhesion to the metal parts without gaps or visible cracks. This ensures that the insulator can withstand current surges during operation and prevents current leakage.<br>(4) The connection between the insulator and the metal parts should have good adhesion and tight fastening. This ensures that the insulator can be stably suspended and prevents current leakage.<br>(5) Good connections between suspension insulators, with complete spring pins and cotter pins. This ensures that the suspension insulators can be stably suspended and prevents current leakage.<br>(6) The insulator body should have good linearity, with a bending radius not exceeding 1%. This ensures that the suspension insulators can be stably suspended and prevents current leakage.<br>(7) The surface of the insulator should be free from obvious discharge marks, circular or through cracks. This ensures that the suspension insulators can be stably suspended and prevents current leakage. |
| LLM | Insulator Maintenance Technical Standards:<br>(1) Insulator Material: The material used for insulators must have good electrical insulation properties and sufficient stability and reliability under different voltages. Commonly used materials include ceramics, glass, and fibers.<br>(2) Insulator Structural Design: Insulators' structures must withstand voltage while maintaining mechanical strength and stability. The design of insulators needs to consider factors such as voltage, frequency, and load, and appropriate materials should be used for manufacturing.<br>(3) Insulator Maintenance: Insulators need regular maintenance to maintain their electrical insulation performance. Insulator maintenance includes cleaning, oiling, and refurbishment. These maintenance tasks can improve the electrical insulation performance of insulators and reduce voltage drops caused by dirt and oil stains on the insulator surface.<br>(4) Insulator Voltage Testing: During insulator maintenance, voltage testing of the insulators is necessary to ensure their electrical insulation performance meets the requirements. The testing voltage is typically between 500V and 1500V, and the appropriate voltage value is selected based on the specific maintenance situation.<br>(5) Insulator Maintenance Records: During insulator maintenance, it is important to record relevant information such as maintenance time, content, and results. This facilitates the evaluation and analysis of the performance of the insulators.<br>(6) The insulator maintenance technical standards aim to ensure that insulators have good electrical insulation performance while enhancing their mechanical strength and stability, thereby guaranteeing their normal service life. |

*B. LKB-E-LLM for Wind Turbines*

This study also constructed a small LKB related to wind turbines to explore the performance of LKB-E-LLM in wind turbines. It mainly includes professional knowledge documents such as "1.5MW Wind Turbine Operation and Maintenance Manual," "Doubly-Fed Wind Turbine Fault Troubleshooting Manual," "Wind Turbine Fault Handling Manual," and "UP-1500 Wind Turbine Maintenance Manual." These documents primarily cover the operation, maintenance, and fault troubleshooting of wind turbines, which are expertise and technical details that most ChatGPT-Like LLMs do not possess. Considering the paper's length limitations, this study presents the results of three professional technical question-and-answer pairs. Then, a comparison is made between the responses of LKB-E-LLM and regular LLM to validate the performance gains brought by the LKB. The detailed question-and-answer records are shown in TABLE II.

As shown in TABLE II, we consulted with LKB-E-LLM and LLM about the repair of wind turbine blades, high temperature of generator carbon brushes, fault code query, and other related technical issues. They behave similarly as they do in the HSTs scenario. The responses from LKB-E-LLM are more professional, detailed, and contain rich technical details relevant to the questions, providing practical guidance. However, due to a lack of contextual information, LLM's responses are severely unrelated to the questions and only offer vague and generic information. For example, in question 3, LKB-E-LLM accurately explains the specific technical details and steps for blade damage repairs, while LLM misunderstands the question and provides a completely incorrect response. Similarly, for question 4, LKB-E-LLM's response contains more professional technical content than LLM, making it valuable guidance for operational and management personnel. Additionally, question 5 validates the fault troubleshooting capability of LKB-E-LLM. LKB-E-LLM can accurately output the relevant fault category and details with the given fault code. In contrast, due to a lack of relevant domain knowledge, LLM provides a response unrelated to wind turbines. Overall, LKB-E-LLM significantly improves the lack of professional knowledge in LLM. This allows LKB-E-LLM to be applied in various professional technical fields, fully leveraging the advantages of LLM and enhancing the effectiveness of industrial PHM-related applications.

TABLE II
LKB-E-LLM AND LLM Q&A RECORDS ON WIND TURBINES RELATED QUESTIONS

| Question-3 | For blade repair, please describe in detail the repair methods for surface and structural damage. |
|---|---|
| LKB-E-LLM | (1) Surface Damage Repair Methods:<br>● Clean the surface with acetone and then use a multi-layer repair material for the repair.<br>● For damaged areas with scratches, grooves, marks, or scrapes on the surface, the fiberglass can be polished into a step shape, increasing by 20mm for each layer. After polishing, clean the surface and then use a multi-layer repair material for the repair.<br>● If there are defects in the seal between the rain cover and the blade casing, it needs to be inspected and repaired.<br>● For corrosion on the blade surface, reinforce it with an epoxy resin hardener.<br>● For structural damage, such as cracks or holes on the blade, reinforce it with epoxy resin hardener or other materials based on the extent of the damage, and then perform polishing and smoothing.<br>(2) Structural Damage Repair Methods:<br>● Inspect the extent of structural layer damage and reinforce it accordingly.<br>● Apply epoxy resin hardener paste to broken glass fiber cloth, then cure and polish it.<br>● For cracks or fissures appearing at the blade root or load-carrying section, immediate shutdown, and repair are required.<br>(3) Maintenance Types:<br>● Routine maintenance: Inspect the blade surface, noise, and lightning protection system, and perform maintenance and record-keeping.<br>● Annual maintenance: Inspect the blade bolts and installation bolts, and perform maintenance and record-keeping.<br>When repairing the blade surface, wearing safety masks and gloves is necessary, as acetone may cause skin irritation.<br>When repairing the blade, reinforcement treatment should be applied to the structural layer to prevent recurring structural damage. Additionally, for surface damage, clean the surface with acetone and then use a multi-layer repair material for the repair. |
| LLM | (1) Surface Damage Repair Methods:<br>● Welding: Welding is a commonly used method for repairing surface damage. During the welding process, methods such as gas-shielded welding, resistance welding, or manual welding can be used. Welding can repair mild surface damage on the blade, for example, by forming a weld seam along the blade's edge to enhance its strength.<br>● Hot pressing: Hot pressing is heating the metal to a high temperature and then applying pressure to repair the damaged area. Hot pressing can repair moderate surface damage on the blade, such as forming a hot pressing rod at the bending section of the blade and applying heat and pressure to strengthen it.<br>(2) Structural Damage Repair Methods:<br>● Laser cutting: Laser cutting is commonly used for repairing structural damage. During the cutting process, CO2 lasers or other laser beams can be used. Laser cutting can repair severe structural damage on the blade, such as cutting at the fracture point of the blade to strengthen it.<br>● Vegetation: Vegetation is a method of implanting plant tissue onto a metal surface for repair. During vegetation, extracts from plant stem or roots can be used to improve the environment of the metal surface and promote plant tissue growth. This method can repair severe structural damage on the blade, such as vegetation at the fracture point of the blade to strengthen it.<br>Surface damage repair methods mainly involve welding, hot pressing, and laser cutting, while structural damage repair methods involve vegetation and metal reinforcement techniques. |
| Question-4 | **Analyze the reasons for the high temperature of the generator carbon brush in detail.** |
| LKB-E-LLM | (1) Overall Current Distribution Imbalance: During generator operation, the working current of each carbon brush should remain constant between 20 to 100A. However, the current may deviate from the overall balance due to carbon brush wear or mechanical faults. If the overall current distribution balance is disrupted, it can lead to an increase in carbon brush temperature.<br>(2) Insufficient Contact Area between Carbon Brush and Slip Ring: Insufficient contact area between the carbon brush and slip ring can result in mechanical wear, increasing carbon brush temperature. During generator maintenance, measuring the contact area between the carbon brush and slip ring is necessary. If it is too small, the carbon brush may need to be replaced or the contact area adjusted.<br>(3) Blockage of Slip Ring Ventilation Holes: Blockage of the slip ring ventilation holes prevents proper airflow, causing the internal heat of the generator to be unable to dissipate. This ultimately leads to an excessive temperature of the carbon brushes. It is important to check for any blockages in the slip ring ventilation holes and promptly clean them if necessary.<br>(4) Excessive Spring Pressure: Excessive spring pressure can cause mechanical wear, increasing carbon brush temperature. Adjusting the spring pressure to ensure it falls within the manufacturer's specified range is necessary.<br>To address the above issues, regular maintenance of the generator is required. This includes monitoring the current distribution balance, the contact area between the carbon brushes and slip rings, and checking for blockage in the slip ring ventilation holes. Any problems should be detected promptly and addressed accordingly. Additionally, regular maintenance of the carbon brushes is essential to ensure their normal operating temperature and lifespan. |
| LLM | (1) Generator Design Issues: There may be certain flaws in the design and manufacturing process of the generator, resulting in uneven or excessive wear on the surface of the carbon brushes. This can affect the heat dissipation efficiency of the carbon brushes, leading to an increase in temperature.<br>(2) Insufficient Number of Carbon Brushes: There may be an insufficient number of carbon brushes, which fails to effectively transfer heat, increasing the surface temperature of the carbon brushes.<br>(3) Quality Issues with Carbon Brushes: There may be quality issues with the carbon brushes, such as impurities in the material or uneven surfaces. This can also impact the heat dissipation efficiency of the carbon brushes.<br>(4) Load Issues during Operation: The generator may experience high loads during operation, increasing the carbon brushes' surface temperature.<br>(5) Improper Maintenance: Improper maintenance of the generator, such as inadequate cleaning of the carbon brushes or failure to replace them in a timely manner, can also cause |
| | an increase in the surface temperature of the carbon brushes. |
| Question-5 | **Fault Code: 050001** |
| LKB-E-LLM | 050001 Fault Code: Hydraulic System Heater Protection Signal Input is 0.<br>This fault code indicates that the protection signal input of the hydraulic system heater is 0. This fault is usually caused by a damaged heater in the hydraulic system or a sensor malfunction, resulting in an erroneous signal input. The shutdown task associated with this fault code is to inspect the condition of the heater in the hydraulic system and carry out repairs or replacement.<br>050000 Fault Code: At least one hydraulic-related fault has occurred.<br>This fault code indicates that there is at least one hydraulic-related fault in the system, requiring further troubleshooting. This typically involves a detailed inspection and testing of the hydraulic system to determine the specific cause of the fault. In some cases, it may be necessary to replace or repair hydraulic components or perform retesting and calibration of the system. |
| LLM | Fault Code 050001 indicates a hardware failure in the television. It typically signifies that the television is experiencing issues with startup or unstable operation.<br>The cause of this fault code is attributed to internal circuitry or hardware problems within the television. This results in the improper transmission of signal inputs to the television's decoder or processor, leading to the inability to start or unstable operation of the television. |

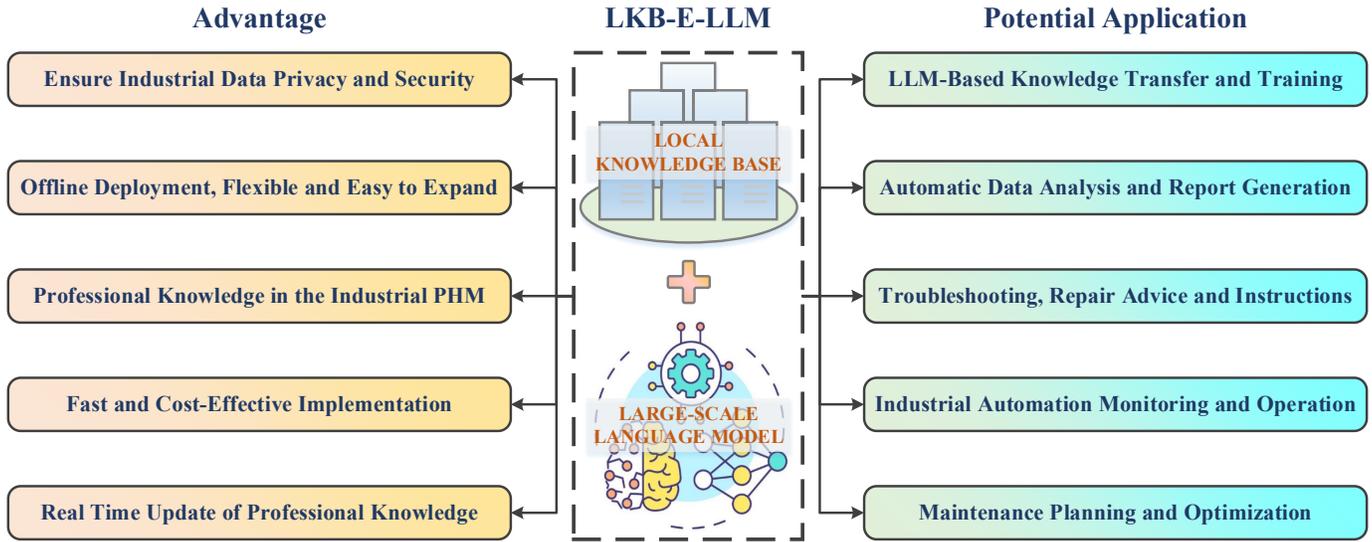

**Fig. 4** Advantages and potential applications of LKB-E-LLM in the field of PHM.

## V. DISCUSSIONS

### A. Advantages of LLM Empowered by LKB

Although ChatGPT-Like LLM exhibits robust language understanding and text generation capabilities, its lack of domain-specific expertise hinders its performance in industrial domains. However, incorporating LKBs into LLM, referred to as LKB-E-LLM, effectively addresses this limitation and significantly enhances the applicability of LLM in the field of industrial PHM.

Fig. 4 illustrates the superior performance of LKB-E-LLM compared to conventional LLM in various aspects. Notably, LKB-E-LLM offers the flexibility of offline deployment and eliminates the need for additional data fine-tuning, thus ensuring the confidentiality and security of sensitive industrial data. Moreover, while LKBs encompass detailed domain-specific knowledge, ChatGPT-Like LLM possesses comprehensive general knowledge and natural language inductive reasoning capabilities. The fusion of these two components facilitates the utilization of the specialized knowledge from the LKB and the vast background knowledge of LLM, thereby enabling the provision of comprehensive, professional, and tailored answers and solutions to users.

In addition, combined with the LKB, the capabilities of LLM can be deeply customized to meet the various needs of industrial PHM. By leveraging a lightweight language model, LKB-E-LLM achieves comparable performance to that of super-large language models, as demonstrated in this study that implements LKB-E-LLM for industrial PHM on consumer-grade computer hardware, delivering excellent performance. Consequently, LKB-E-LLM presents a cost-effective and rapid implementation solution suitable for deployment in resource-constrained industrial scenarios.

Lastly, given the dynamic nature of professional knowledge and documents in the industrial domain, it is imperative for ChatGPT-Like LLM to possess real-time knowledge updating capabilities. The LKB-based solution effectively fulfills this requirement, as the LKB can be dynamically updated without necessitating modifications to LLM. This ensures that LKB-E-LLM remains up-to-date with the latest technical knowledge, aligning with the evolving demands of industrial PHM applications.

In conclusion, the integration of LKB and ChatGPT-Like LLM in specific domains capitalizes on their respective strengths, offering comprehensive, accurate, professional, and personalized information services to meet the specific needs of the industrial PHM field.

### B. Potential Applications of LLM Empowered by LKB

Although ChatGPT-Like LLM mainly uses text dialogue as the main application scenario, LLM combined with LKB will significantly enhance its application potential in industrial PHM and have a revolutionary impact on the field of PHM. Fig. 4 illustrates the comprehensive advantages of LKB-E-LLM across various dimensions within the industrial PHM domain, including enhanced efficiency, revenue generation, performance improvement, and cost reduction. For instance,

LKB-E-LLM facilitates LLM-based knowledge dissemination and training, enabling industrial enterprises to swiftly cultivate competent and dependable personnel, thereby reducing the time and cost associated with training and maintaining operational staff. Moreover, LKB-E-LLM autonomously analyzes industrial data to generate real-time analysis reports, aiding engineers in monitoring industrial operations and offering troubleshooting guidance and maintenance recommendations. Notably, LKB contains detailed descriptions of equipment and system failure scenarios, along with optimal solutions. Leveraging LKB-E-LLM, junior engineers can promptly identify faults and determine the most suitable solutions, consequently enhancing the efficiency of industrial PHM and diminishing maintenance costs. Furthermore, extensive repositories of maintenance plans and measures have accumulated over the long-term operation of industrial enterprises. LKB-E-LLM can analyze and summarize these cases, facilitating the generation of appropriate maintenance plans based on current equipment and system conditions, as well as optimizing existing plans, thereby augmenting the benefits derived from maintenance activities.

Through these advancements, LKB-E-LLM actively engages in industrial operation and maintenance, contributing to the realization of automated industrial processes and improving the outcomes of PHM applications. Consequently, research on LLM-based PHM applications is progressing rapidly, with the potential to revolutionize the established paradigms within the field in the foreseeable future.

*C. Future Directions of LLM Empowered by LKB*

It can be seen from Section III that the realization of LKB-E-LLM mainly relies on the construction of domain-specific LKB, the representation ability of the text embedding model, and the language processing, induction, and reasoning capabilities of LLM. Future advancements can be explored and enhanced within these three directions. The accuracy, comprehensiveness, and professionalism of the LKB play a crucial role in enabling LKB-E-LLM to address the diverse demands of industrial PHM applications. Hence, industrial enterprises must maintain a complete offline LKB, ensuring timely updates and additions of the latest professional knowledge and data. The effectiveness of the text embedding model determines the accurate representation of information in the knowledge base, directly impacting the ability to search for query-related knowledge accurately. Consequently, further investigation into cutting-edge and novel text embedding models will enhance the performance of LKB-E-LLM. ChatGPT-Like LLM serves as the foundation of LKB-E-LLM, and its language processing capability directly influences the ability of engineers to obtain precise and concise results. Considering that a significant portion of data in the industrial PHM field exists in the form of videos and images, the exploration of a reliable visual language model [42] and its integration into LKB-E-LLM can significantly enhance its application potential.

## VI. CONCLUSIONS

PHM is crucial in ensuring industrial systems and equipment's reliable and safe operation. With significant advancements in recent years, ChatGPT-Like LLM has demonstrated remarkable progress in language understanding, text generation, and generalization, offering the potential to transform the application landscape of industrial PHM and facilitate the development of highly intelligent industrial operation and maintenance systems. However, the lack of domain-specific expertise poses a challenge in effectively applying ChatGPT-Like LLM to address professional PHM problems. Therefore, this study focuses on exploring the solution of LKB-based LLM for industrial PHM, aiming to overcome the aforementioned limitations. The study provides comprehensive details on the methodology and technical aspects of the proposed solution. Through two real-world PHM case studies, it is demonstrated that integrating LLM with LKB can better cater to the requirements of industrial PHM applications. Compared to traditional LLM approaches, the combined LLM and LKB approach enables the acquisition of extensive domain-specific knowledge in industrial PHM, with the ability to update knowledge in real-time, while also offering advantages such as rapid and cost-effective deployment. Consequently, LLM combined with LKB exhibits significant application potential in industrial PHM, with the potential to revolutionize the PHM domain.